\documentclass{webofc}
\usepackage[utf8]{inputenc}
\usepackage{graphicx}
\usepackage[varg]{txfonts}
\usepackage{subcaption}
\usepackage{appendix}
\usepackage[capitalise]{cleveref}
\usepackage{amsmath}
\usepackage{amssymb}
\usepackage{float}
\usepackage{bm}
\usepackage{xcolor}

\newcommand{\onedot}{.\\}

\newcommand{\etal}{\emph{et al}\onedot}

\newcommand{\vx}{\ensuremath{\mathbf{x}}}
\newcommand{\vz}{\ensuremath{\mathbf{z}}}
\newcommand{\vy}{\ensuremath{\mathbf{y}}}

\newcommand{\f}{f}

\newcommand{\dsep}{\ensuremath{\;\|\;}}

\newcommand{\pdata}{\ensuremath{p_{\text{data}}}}
\newcommand{\pfake}{\ensuremath{ p_{G} }}

\newcommand{\E}{\ensuremath{\mathbb{E}}}

\newcommand{\geant}{\texttt{Geant4} }

\begin{document}

\title{Generative Models for Fast Calorimeter Simulation: the LHCb case}

\author{
 \firstname{Viktoria} \lastname{Chekalina} \inst{1,2}
\and
    \firstname{Elena} \lastname{Orlova} \inst{3}
\and
    \firstname{Fedor} \lastname{Ratnikov} \inst{1,2}\fnsep\thanks{\email{fedor.ratnikov@cern.ch}}
\and
     \firstname{Dmitry} \lastname{Ulyanov} \inst{3}
\and
     \firstname{Andrey} \lastname{Ustyuzhanin} \inst{1,2}
\and
     \firstname{Egor} \lastname{Zakharov} \inst{3}
}

\institute{
NRU Higher School of Economics, Moscow, Russia
\and
Yandex School of Data Analysis, Moscow, Russia  
\and
Skolkovo Institute of Science and Technology, Moscow, Russia
}
        
\abstract{Simulation is one of the key components in high energy physics. Historically it relies on the Monte Carlo methods which require a tremendous amount of computation resources. These methods may have difficulties with the expected High Luminosity Large Hadron Collider (HL-LHC) needs, so the experiments are in urgent need of new fast simulation techniques. We introduce a new Deep Learning framework based on Generative Adversarial Networks which can be faster than traditional simulation methods by 5 orders of magnitude with reasonable simulation accuracy. This approach will allow physicists to produce a sufficient amount of simulated data needed by the next HL-LHC experiments using limited computing resources.}

\maketitle

\section{Introduction}

Simulation plays an important role in particle and nuclear physics. It is widely used in detector design and in comparisons between experimental data and theoretical models. Traditionally, simulation relies on \textit{Monte Carlo methods} and requires significant computational resources. In particular, such methods do not scale to meet the growing demands resulting from large quantities of data expected during High Luminosity Large Hadron Collider (HL-LHC) runs. The detailed simulation of particle collisions and interactions as captured by detectors at the LHC using a well-known simulation software \geant annually requires billions of CPU hours constituting more than half of the LHC experiments' computing resources~\cite{bozzi2014,flynn2015computing}. More specifically, the detailed simulation of particle showers in calorimeters is the most computationally demanding step.
 
A line of simulation methods that exploit the idea of reusing previously calculated or measured physical quantities have been developed to reduce the computation time~\cite{grindhammer2000parameterized,atlas2010simulation}. These approaches suffer from being specific to an individual experiment and, despite being faster than the full simulation, they are not fast enough or lack accuracy. Thus, the particle physics community is in need of new faster simulation methods to model experiments. 
    
One of the possible approaches to simulate the calorimeter response is using \textit{deep learning} techniques. In particular, a recent work~\cite{paganini2017calogan}, provided evidence that \textit{Generative Adversarial Networks} can be used to efficiently simulate particle showers. While over $100,000 \times$ speed-up over \geant is achieved, the setup was quite simple as the input particles were parametrized by energy only. However,  even in this simplified approach, there are significant differences in distributions between generated and original parameters. 

In this work we build a model upon \text{Wasserstein Generative Adversarial Networks} and show its superior performance over approach~\cite{paganini2017calogan}. We also evaluate our model in a more complex scenario, when a particle is described by $5$ parameters: 3d momentum $(p_x,~ p_y,~ p_z)$ and 2d coordinate $(x,~ y)$. Our method for high-fidelity fast simulation of particle showers in the specific LHCb calorimeter aims to replace the existing Monte Carlo based methods and achieve a significant speed-up factor.

\section{Related work: GANs basics and GANs in HEP}
Generative models are of great interest in deep learning. With these models, one can approximate a very complex distribution defined as a set of samples. 
For example, such models can be utilized to generate a face image of a non-existing person or to continue a video sequence given several initial frames. 
In this section, we give a brief overview of the most popular generative model in computer vision — Generative Adversarial Networks (GANs),
 its strong and weak sides and different modifications to alleviate its weaknesses. Then, we review and analyse current approaches for applying GANs to the simulation of calorimeters in High energy physics.

\subsection{Background: from GAN to conditional WGAN}

Generative Adversarial Networks (GANs) were originally presented by I.~Goodfellow~\etal in 2014 \cite{goodfellow2014generative} and quickly became a state-of-the-art technique in areas such as image generation \cite{radford2015unsupervised}, with a huge number of extensions \cite{IsolaZZE16,CycleGAN2017,wang2018video}.

In the GAN framework, the aim is to learn a mapping $G$, usually called \textit{generator}, to warp an easy-to-draw distribution $p(\vz)$ (e.g. $p(\vz) = \mathcal{N}(0, I)$) into a target distribution $\pdata(\vx)$ to facilitate sampling from $\pdata(\vx)$. When $G$ is learned, $G \equiv G^*$, sampling from the target distribution $\pdata(\vx)$ is done by first drawing a sample from the distribution $p(\vz)$ and then feeding the sample into the generator: $G^*(\vz) \sim \pdata$, where $\vz \sim p(\vz)$. For such sampling procedure, the time needed to draw a sample from $\pdata(\vx)$ is approximately equal to the time needed to evaluate the function $G$ in a point.  

The generator is learned by using a feedback from an external classifier (usually called \textit{discriminator}), which tries to find discrepancy between the target distribution $\pdata(\vx)$ and fake distribution $\pfake(\vx)$ defined by samples from the generator $G(\vz) \sim \pfake(\vx),\, \vz \sim p(\vz)$. %The process in summarised in~\cref{fig:GANs}.

% \begin{figure}
% \centering
% \includegraphics[width=0.3\linewidth]{figures/gan_pic.pdf}
% \caption{Generative Adversarial Networks for digit generation. The generator $G$ transforms the noise vector $\vz \sim p(z)$ to an image of a digit and the discriminator $D$ classifies inputs as real digits or fake digits from generator. Generator and discriminator are trained in an adversarial manner: the task of $G$ is to make it impossible for $D$ to distinguish between the real and fake digits as in this case $G$ reproduces the data distribution $\pdata$.}\label{fig:GANs}
% \end{figure}

More formally, generator $G$ and discriminator $D$ play the following zero sum game: 
\begin{equation}\label{eq:gan}
\min_G \max_D \E_{\vx \sim \pdata(\vx)} [\log D(\vx)] + \E_{\vx \sim \pfake(\vx)} [\log(1 - D(\vx))]\, ,
\end{equation} 
where $D(G(\vz))$ is the output of the discriminator specifying the probability of its input to come from the target distribution $\pdata$.

In practice, the mappings $G$ and $D$ are parametrized by deep neural networks and the objective~\cref{eq:gan} is optimized using alternating gradient descent. For a fixed generator, the discriminator minimizes binary cross-entropy in a binary classification problem (samples from $\pdata$ versus samples from $\pfake$). For the fixed discriminator, the generator is updated to make its samples to be misclassified by the discriminator, thus moving the  fake distribution closer to the target distribution.   

For a fixed generator, it is possible to show that  the optimal value for the inner optimization can be written analytically: 
\begin{equation}\label{eq:js}
\max_D \E_{\vx \sim \pdata(\vx)} [\log D(\vx)] + \E_{\vx \sim \pfake(\vx)} [\log(1 - D(\vx))] = \text{JS}( \pdata \dsep \pfake)\, , 
\end{equation} 
where $\text{JS}$ is the Jensen-Shannon divergence. In fact, for the fixed generator (hence fixed fake distribution), the discriminator computes the divergence between the target distribution $\pdata$ and the fake distribution $\pfake$. When the divergence is computed, the generator aims to update the fake distribution to make this divergence lower: $\min_G \text{JS}( \pdata \dsep \pfake )$. While the Jensen-Shannon divergence naturally arises from the original game~\cref{eq:gan}, any divergence or distance $\mathcal{D}$ can be used instead: $\min_G \mathcal{D}( \pdata \dsep \pfake )$.
A recent work~\cite{arjovsky2017wasserstein} proposed to use the \textit{Wasserstein distance} instead of the Jensen-Shannon divergence proving its better behavior:
\begin{equation}\label{eq:wasserstein_metric}
W(\pdata \dsep \pfake ) = \max_{\f \in \mathcal{F}} \E_{\vx \sim \pdata(\vx)}[\f(\vx)] - \E_{\vx \sim \pfake} [\f(\vx)]
\end{equation}
where $\mathcal{F}$ is a set of 1-Lipshitz functions. Using the Wasserstein distance instead of the Jensen-Shannon divergence in the GAN objective leads to the Wasserstein GAN (WGAN) objective: 
\begin{equation}\label{eq:wgan}
\min_G \max_{\f \in \mathcal{F}} \E_{\vx \sim \pdata(\vx)}[\f(\vx)] - \E_{\vx \sim \pfake(\vx)} [\f(\vx)]\, .
\end{equation}

It is highly non-trivial to search over the set of 1-Lipshitz functions and several ways have been proposed in order to force this constraint \cite{arjovsky2017wasserstein,gulrajani2017improved}. In Ref. \cite{gulrajani2017improved}, it is proved that the set of optimal functions for~\cref{eq:wgan} contains such function, that the norm of it's gradient in any point equals one. In practice, this result motivates an additional loss added to the objective~\cref{eq:wgan} with a weight $\lambda$, while the hard constraint on the function $\f$ to belong to the set $\mathcal{F}$ is removed and $\f$ is searched over all possible functions:    
\begin{equation}\label{gpwgan-loss}
% \begin{gathered}
\min_G \max_\f \E_{\vx \sim \pdata(\vx)}  \f(\vx) - \E_{\vx \sim \pfake(\vx)} \f(\vx) + 
\lambda \E_{\vx \sim \pfake} \big(\|\nabla_{\tilde{\vx}} D(\tilde{\vx})\|_2 - 1\big)^2 .
% \end{gathered}
\end{equation}

WGAN can be easily adapted to model a conditional distribution $\pdata(\vx | \vy)$. The generator is modified to take the condition along with the sample $\vz$ so the fake distribution is now defined as $G(\vz, \vy) \sim \pfake(\vx | \vy),\, \vz \sim p(\vz)$ and the game is 
\begin{equation}\label{CWGAN}
% \begin{gathered}
\min_G \max_\f \E_{\vy \sim p(\vy)} \Big[ \E_{\vx \sim \pdata(\vx | \vy)}  \f(\vx) - \E_{\vx \sim \pfake(\vx | \vy)} \f(\vx) + 
\lambda \E_{\vx \sim \pfake(\vx | \vy)} \big(\|\nabla_{\tilde{\vx}} D(\tilde{\vx})\|_2 - 1\big)^2 \Big]\,.
% \end{gathered}
\end{equation}

\subsection{GANs in high energy physics}
A  systematic study on the application of deep learning to the simulation of calorimeters for particle physics has been carried out by Paganini et al. in 2017~\cite{paganini2017calogan} and has resulted in the CaloGAN package. The authors aim to speed up particle simulation in a 3-layer heterogeneous calorimeter using GANs framework and achieve $\sim \times 10^5 $ speedup. They used an existing state-of-the-art but slow simulation engine \geant to create a training dataset. They simulated positrons, photons and charged pions with various energies sampled from a flat distribution between 1 GeV and 100 GeV. All incident particles in this study have an initial momentum perpendicular to the face of the calorimeter. The shower in the first layer is represented as a $3 \times 96$ pixel image, the middle layer as a $12 \times 12$ pixel image, and the last layer as a $12 \times 6$ pixel image. 

Their design of the generator network is based on a DCGAN structure \cite{radford2015unsupervised} with some convolutional layers replaced by locally-connected layers \cite{taigman2014deepface}. The idea of locally connected layers is based on the fact that every pixel position gets its own filter while an ordinary convolutional layer is applied over the whole image, independently of location. An extension of this method to particle physics simulation has been described in the previous work of the authors, where the resulting type of neural network was called LAGAN \cite{de2017learning}. A special section in the paper is devoted to the evaluation of the quality of the CaloGAN produced images, where the sparsity level,  energy per layer or total energy, are used as measures of the performance of the model. 

The obtained results demonstrate a prospect of application of GANs for the particle showers generation and its replacement of the Monte Carlo methods with the proposed approach. The CaloGAN approach yields sizeable simulation-time speedups compared to \geant. 

%In fact, the CaloGan model is based on DCGAN with the described tricks. However, GANs tend to suffer from mode collapse. Therefore, the CaloGan architecture cannot be applied for all datasets, because %here is a high probability of mode collapse appearance and it is a limitation of this work.

\section{Dataset}
In this work, we focused on electrons interactions inside an electromagnetic calorimeter inspired by the LHCb detector at the CERN LHC \cite{Alves:2008zz}. The calorimeter in this study uses "shashlik" technology of alternating scintillating tiles and lead plates. The prototype  consists of 5 $\times$ 5 blocks of size 12 cm $\times$ 12 cm, the cell granularity corresponds to each block being 6 $\times$ 6 of size 2 cm $\times$ 2 cm. There are 66 total layers in ECAL, 2 mm lead absorber and 4 mm scintillator each. In fact, the shower appears in 3d, but all energies deposited in all scintillator layers of one cell are summed up. This procedure reproduced the actual shower energy collection in the calorimeter. Thus, the calorimeter response can be represented as 30 $\times$ 30 images $Y$ with the corresponding parameters $(p_x,~ p_y,~ p_z,~ x,~ y)$ of the original particle. An example of such an image is presented in the top row of~\cref{fig:geant_vs_ours}.

The training data set is created as follows. The calorimeter prototype structure described above is described in \geant as a  mixture of subsequent sensitive and insensitive volumes. Particles are generated using a particle gun. Particle energies are distributed dropping as $1/E$ in the energy range between 1 and 100 GeV. Particle positions are generated uniformly in the square 1$\times$1 cm in the centre of the calorimeter face. Finally, particle angles are distributed normally with widths of 20 degrees in $XZ$ plane and 10 degrees in $YZ$ plane. Then \geant is used to simulate particle interaction with the calorimeter using the full set of corresponding physics processes. Information about every event, therefore, includes the original particle parameters accompanied by 30 $\times$ 30 matrix of energies deposited in scintillators for every cell tower $Y$. Electrons are used as test particles. Produced training dataset contains 50 000 events, and another 10 000 events are used as a  test data sample.

\section{Our GAN model} \label{sec:model}
Our idea is to treat simulations as a black-box and replace the traditional Monte Carlo simulation with a method based on Generative Adversarial Networks. As WGANs with gradient penalty are considered to be the state-of-the-art technique for image producing, we implement a tool based on this approach. For it to be useful in realistic physics applications, such a system needs to be able to accept requests for the generation of showers originating from incoming particle parameters such as 3d momentum and 2d coordinate. We introduce an auxiliary task of reconstructing these parameters $p_x$, $p_y$, $p_z$ and $x$, $y$ from a shower image.

\subsection{Model architecture}

We need to generate a specific calorimeter response for a particle with some parameters. It means that the model is required to be conditional.
% sampling not just from $p(\textbf{y}),$ but from $p(\textbf{y}|\vx),$ so, 
Firstly, we describe a generator and discriminator architecture. The generator maps from an input (a 512 $\times$ 1 vector sampled from a Gaussian distribution and the particle parameters) to a 30 $\times$ 30 image $\hat{\textbf{y}}$ using deconvolutional layers (in fact, it is an upsampling procedure and convolutions) which are arranged as follows. We concatenate the noise vector and the parameters $(p_x,~ p_y,~ p_z,~ x,~ y)$, after which we add a fully connected layer with reshaping and obtain a 256 $\times$ 4 $\times$ 4 output. After a sequence of 2d deconvolutions, we get outputs of size  128 $\times$ 8 $\times$ 8, 64 $\times$ 15 $\times$ 16 and 32 $\times$ 32 $\times$ 32  with ReLu activation functions. After this procedure, we crop the last output to obtain the image of the desired size 30 $\times$ 30.

\begin{figure}
\centering
\includegraphics[width=0.95\textwidth]{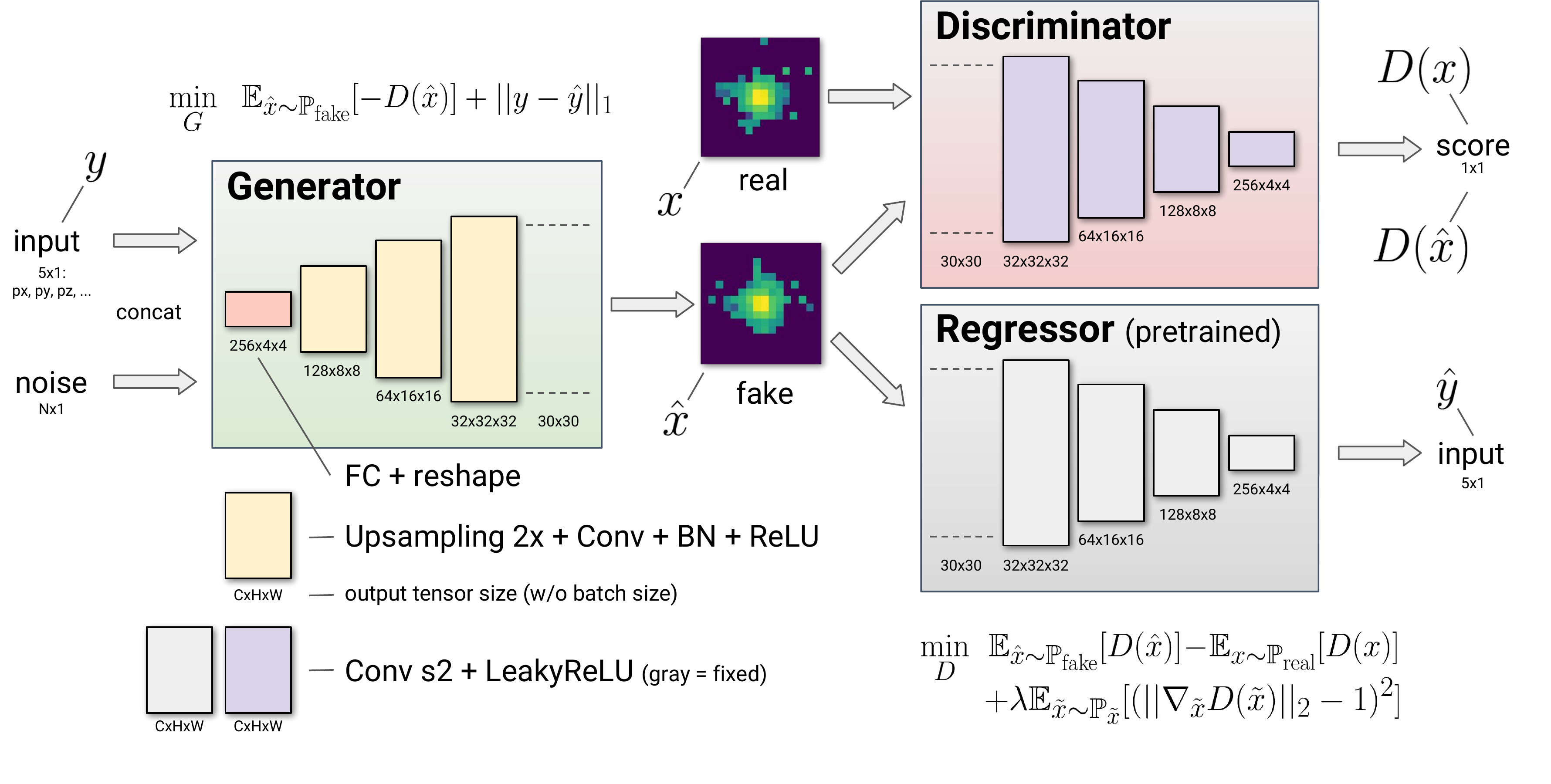}
\caption{Model architecture. Pre-trained regressor for the particle parameters prediction makes our model conditional. Thanks to building up the information from the pre-trained regressor into the discriminator gradient we learn $G$ to produce a specific calorimeter response.}\label{fig:model}
\end{figure}

As for the discriminator, it takes a batch of images as input (all images in the batch are real or generated by $G$) and returns the score $D(\textbf{y})$ or $D(\hat{\textbf{y}})$ as it is described in \cite{arjovsky2017wasserstein}. The discriminator architecture is simply the reversed generator architecture (i.e. sizes of layers go in the opposite order). It implies that we have a 30 $\times$ 30 matrix as input, from which we obtain output layers of size 32 $\times$ 32 $\times$ 32, 64 $\times$ 15 $\times$ 16, 128 $\times$ 8 $\times$  8, followed by reshaping, which  leads to 256 $\times$ 4 $\times$ 4, and by applying LeakyRelu activation function we get the final score. The model scheme is presented in~\cref{fig:model}.

How to train WGAN with gradient penalty in a conditional manner is described in the following section.

\subsection{Training strategy} \label{sec:training_strategy}
Due to the nature of WGAN loss, conditioning on the continuous value is a non-trivial task. To overcome this issue we suggest embedding a pre-trained regressor in our model. We train a neural network to predict the particle parameters by the calorimeter response. As for architecture, it has the same one as the discriminator but with a perceptual loss described in \cite{johnson2016perceptual}, because it was seen to work better compared to standard MSE. By building up the information from the pre-trained regressor into the discriminator gradient, we obtain the conditional model because we train the generator and the discriminator together. As a result, the discriminator makes the generator produce a specific calorimeter response.

\begin{figure}
\begin{center}
\includegraphics[width=0.3\textwidth]{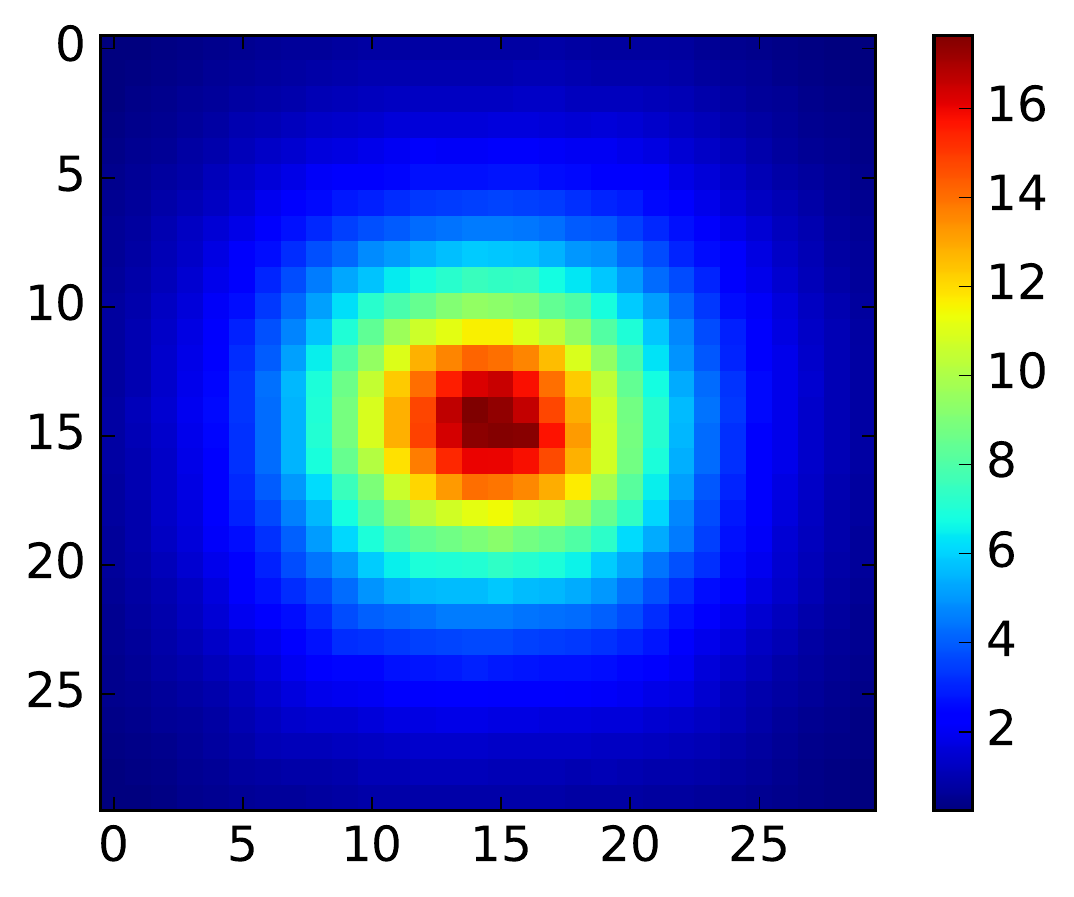}
\caption{energy deposition in different cells of used 30$\times $30 setup for \geant simulated events averaged over all events in the used dataset. \label{fig:real-imgs}}
\end{center}
\end{figure}

Matrices from our dataset are pretty sparse because almost all information is located in central cells (see~\cref{fig:real-imgs}). To make the optimization process easier we apply a box-cox transformation. This mapping helps to smooth the data that makes the optimization process more stable.
Results obtained with the described model are presented in the following section.

%PyTorch \cite{pytorch} library was used to carry out model training and experiments.

\section{Experiments}

\begin{figure}
\captionsetup[subfigure]{justification=centering}
  \centering
  \begin{subfigure}{0.24\textwidth}
    \centering
    \includegraphics[width=1\textwidth]{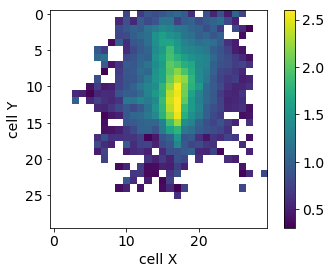}
    % \caption{\\$E = 63.7~\text{GeV}$ \\ $\frac{p_x}{p_z}=0.005$ \\ $\frac{p_y}{p_z}=0.154$}\label{fig:real-imgs-1}
  \end{subfigure}
  \begin{subfigure}{0.24\textwidth}
    \centering
    \includegraphics[width=1\textwidth]{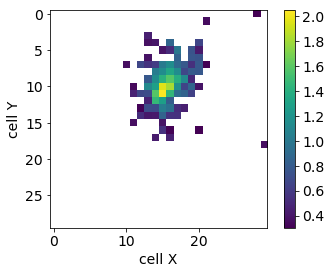}
    % \caption{\\$E = 6.5~\text{GeV}$ \\  $\frac{p_x}{p_z}=0.0046$ \\$\frac{p_y}{p_z}=0.108$}\label{fig:real-imgs-2}
  \end{subfigure}
    \begin{subfigure}{0.24\textwidth}
    \centering
    \includegraphics[width=1\textwidth]{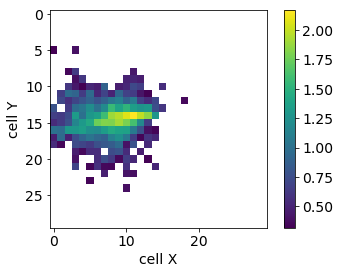}
    % \caption{\\$E = 15.6~\text{GeV}$ \\ $\frac{p_x}{p_z}=0.196$ \\ $\frac{p_y}{p_z}=-0.036$}\label{fig:real-imgs-3}
  \end{subfigure}
  \begin{subfigure}{0.24\textwidth}
    \centering
    \includegraphics[width=1\textwidth]{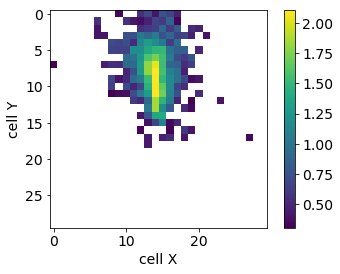}
    % \caption{\\$E = 15.6~\text{GeV}$ \\  $\frac{p_x}{p_z}=-0.019$ \\ $\frac{p_y}{p_z}=0.181$}\label{fig:real-imgs-4}
  \end{subfigure}\\
   \begin{subfigure}{0.24\textwidth}
    \centering
    \includegraphics[width=1\textwidth]{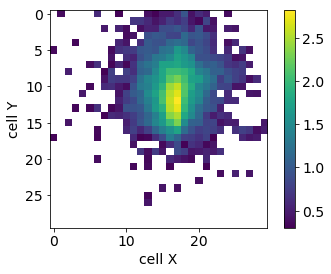}
    \caption{\\$E_0 = 63.7~\text{GeV}$ }%\\ $\frac{p_x}{p_z}=0.005$ ,  $\frac{p_y}{p_z}=0.154$}}%\label{fig:gen-imgs-1}
  \end{subfigure}
  \begin{subfigure}{0.24\textwidth}
    \centering
    \includegraphics[width=1\textwidth]{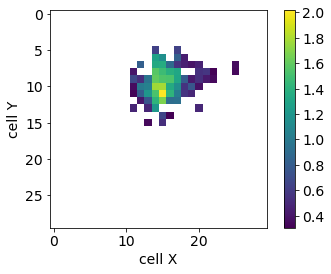}
    \caption{\\$E_0 = 6.5~\text{GeV}$ }% \\  $\frac{p_x}{p_z}=0.046$ , $\frac{p_y}{p_z}=0.108$}}%\label{fig:gen-imgs-2}
  \end{subfigure}
    \begin{subfigure}{0.24\textwidth}
    \centering
    \includegraphics[width=1\textwidth]{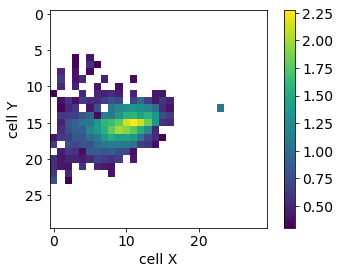}
    \caption{\\$E_0 = 15.6~\text{GeV}$ }% \\ $\frac{p_x}{p_z}=0.196$ ,  $\frac{p_y}{p_z}=-0.036$}}%\label{fig:gen-imgs-3}
  \end{subfigure}
  \begin{subfigure}{0.24\textwidth}
    \centering
    \includegraphics[width=1\textwidth]{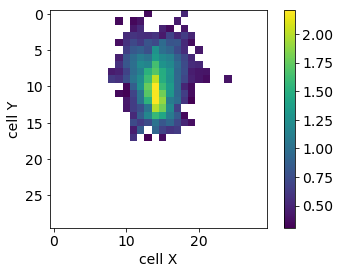}
    \caption{\\$E_0 = 15.9~\text{GeV}$ }% \\  $\frac{p_x}{p_z}=-0.019$ ,  $\frac{p_y}{p_z}=0.181$}}%\label{fig:gen-imgs-4}
  \end{subfigure}
 
  \caption{Showers generated with \geant (first row) and the showers,
    simulated with our model (second row) for three different sets of
    input parameters. Color represents $log_{10}(\frac{E}{MeV})$ for every cell.}
  \label{fig:geant_vs_ours}
\end{figure}

We start with comparing original clusters, produced by full \geant
simulation and clusters generated by the trained model for the same parameters of the incident particles: the same energy, the same direction,
 and the same position on the calorimeter face. Corresponding images
 for four arbitrary parameter sets are presented
 in~\cref{fig:geant_vs_ours}. These images demonstrate the very good visual similarity between simulated and generated clusters.

\begin{figure}
  \centering
  \begin{subfigure}[t]{0.3\textwidth}
    \centering
    \includegraphics[width=1\textwidth]{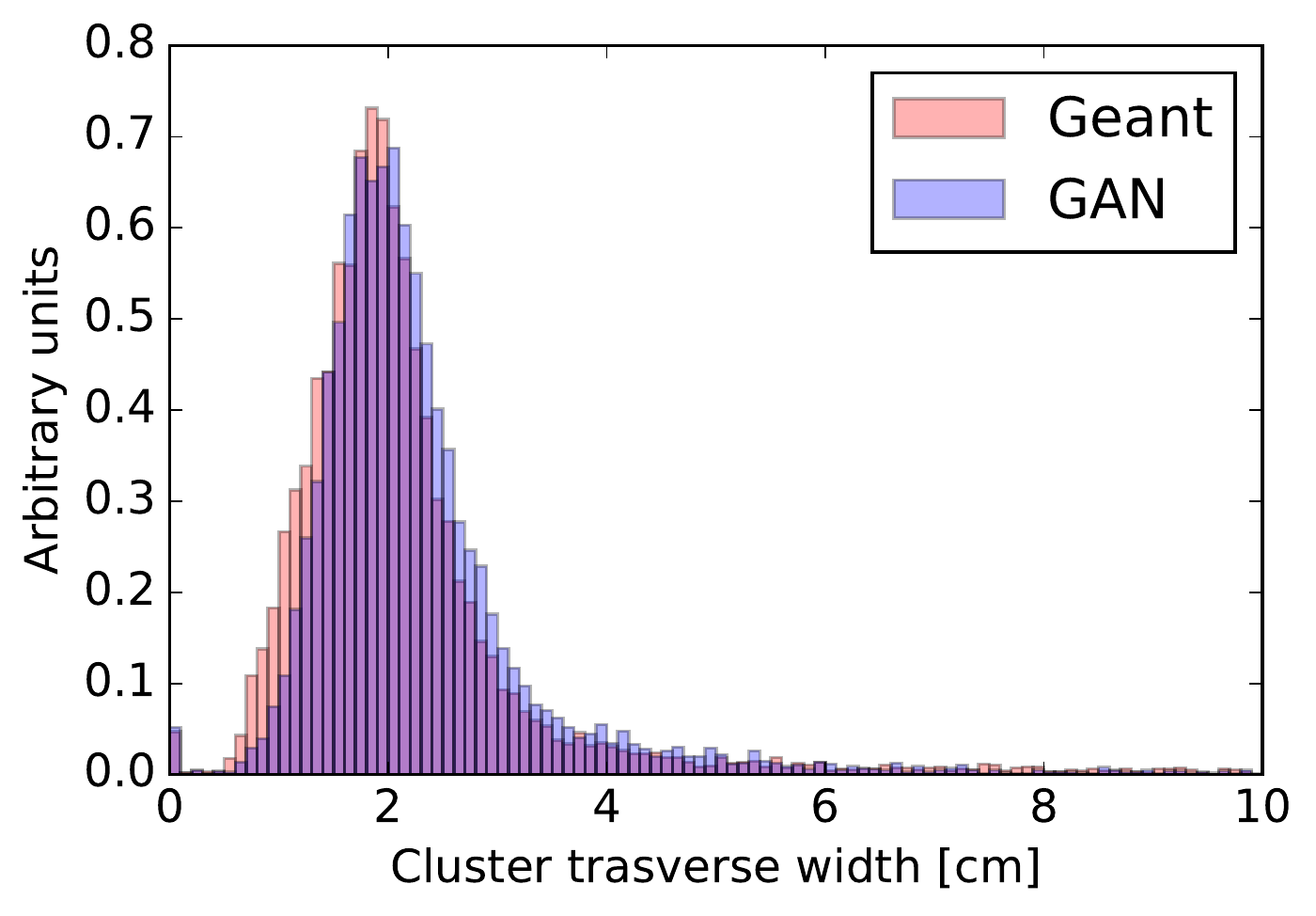}
    \caption{The transverse width of real and generated clusters}
  \end{subfigure}\hspace{0.2\textwidth}
 \begin{subfigure}[t]{0.3\textwidth}
    \centering
    \includegraphics[width=1\textwidth]{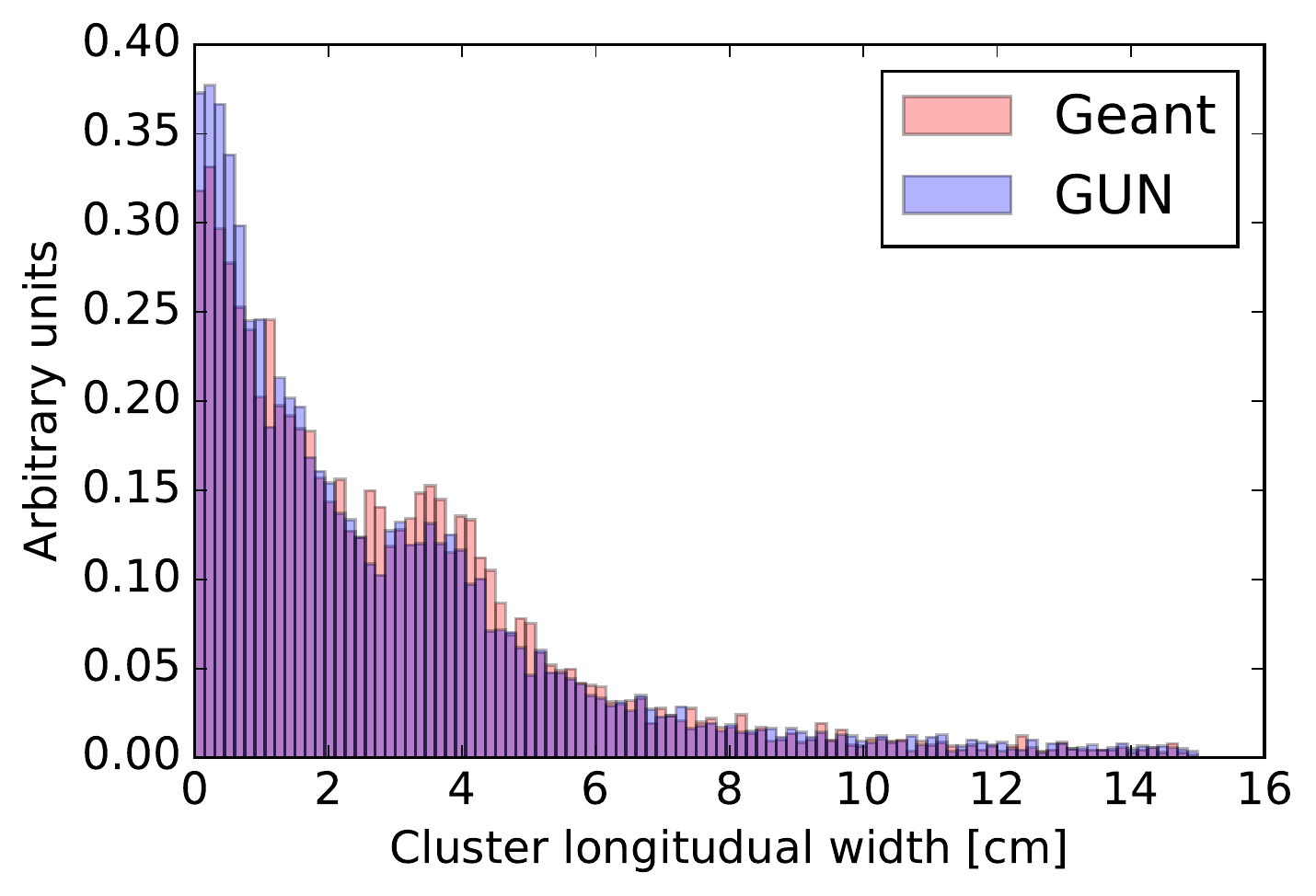}
    \caption{The longitudinal width of real and generated clusters}
  \end{subfigure}
  \begin{subfigure}[t]{0.3\textwidth}
    \centering
    \includegraphics[width=1\textwidth]{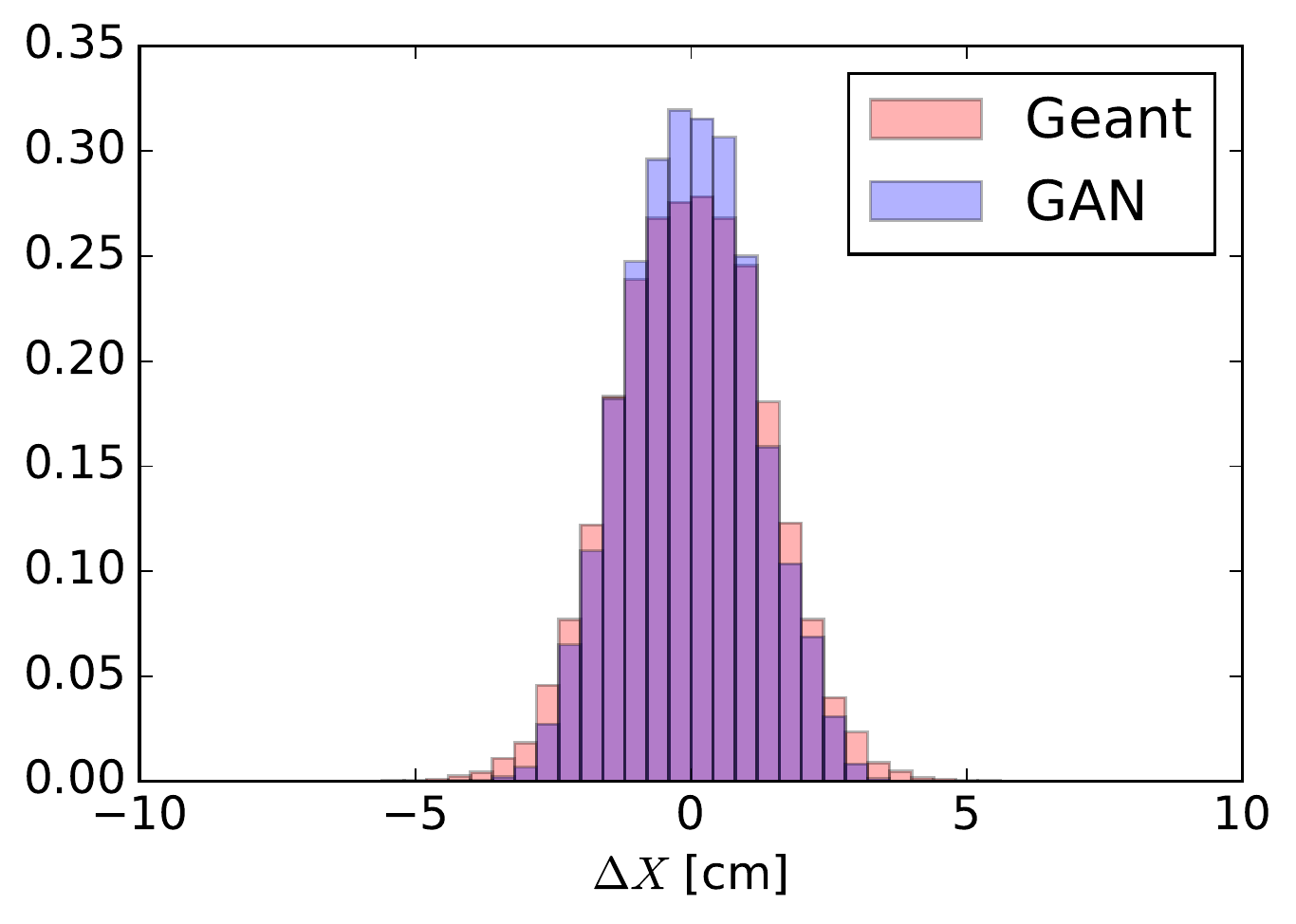}
    \caption{$\Delta X$ between cluster center of mass and the true particle coordinate}
  \end{subfigure}\hspace{0.2\textwidth}
  \begin{subfigure}[t]{0.3\textwidth}
    \centering
    \includegraphics[width=1\textwidth]{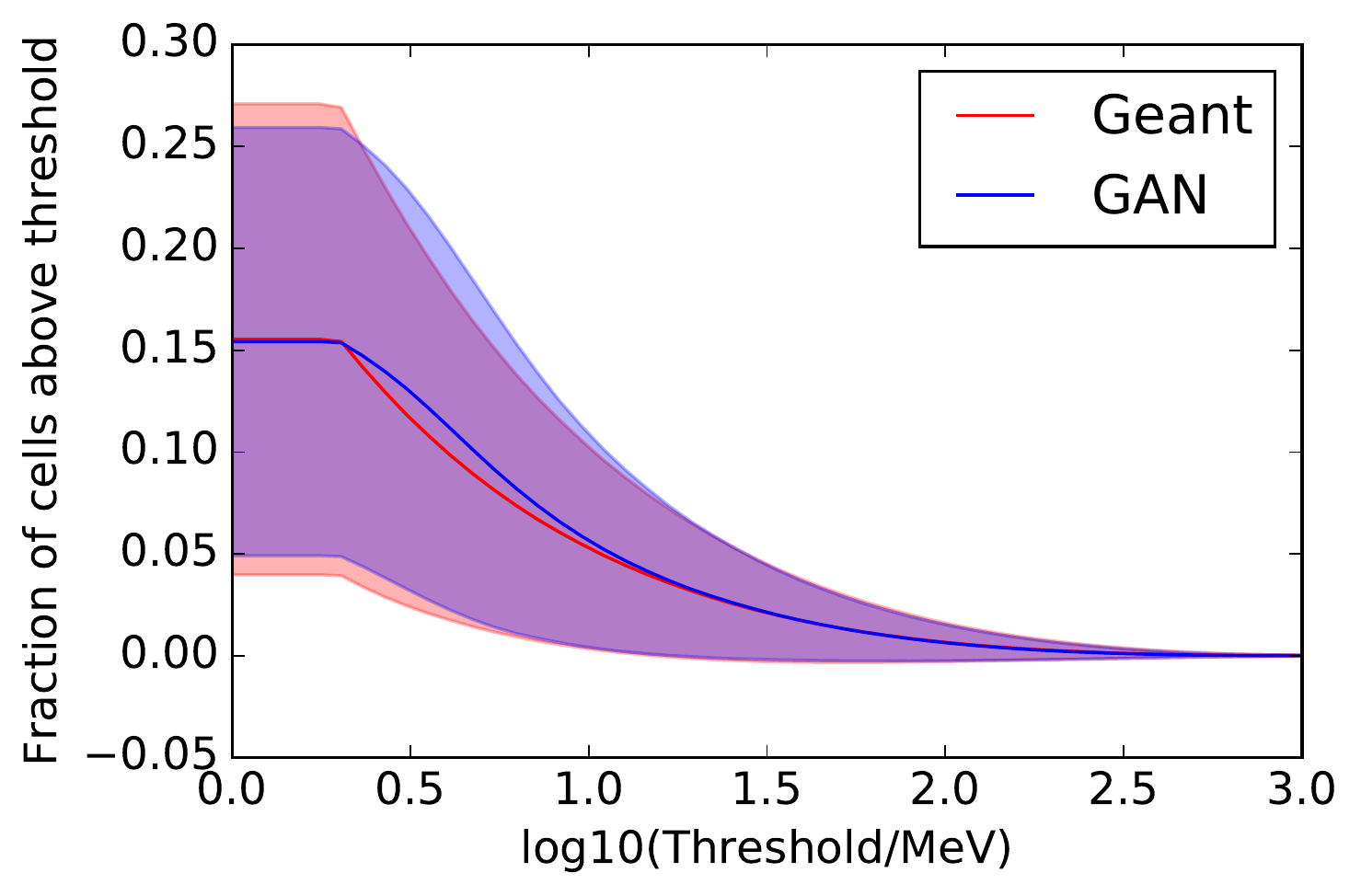}
    \caption{The sparsity of real and generated clusters}
  \end{subfigure}
  \begin{subfigure}[t]{0.3\textwidth}
    \centering
    \includegraphics[width=1\textwidth]{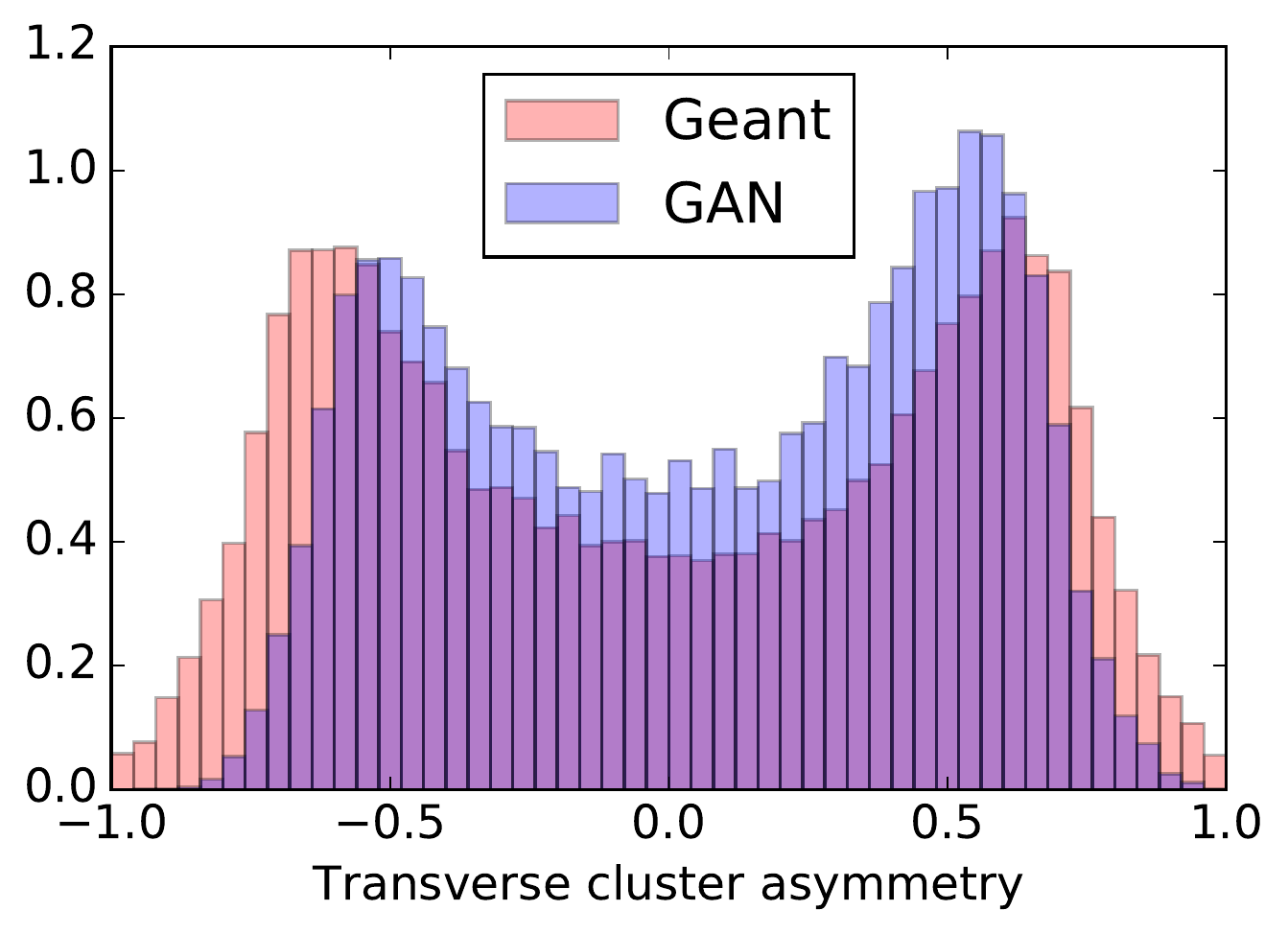}
    \caption{The transverse asymmetry of real and generated clusters}
  \end{subfigure}\hspace{0.2\textwidth}
  \begin{subfigure}[t]{0.3\textwidth}
    \centering
    \includegraphics[width=1\textwidth]{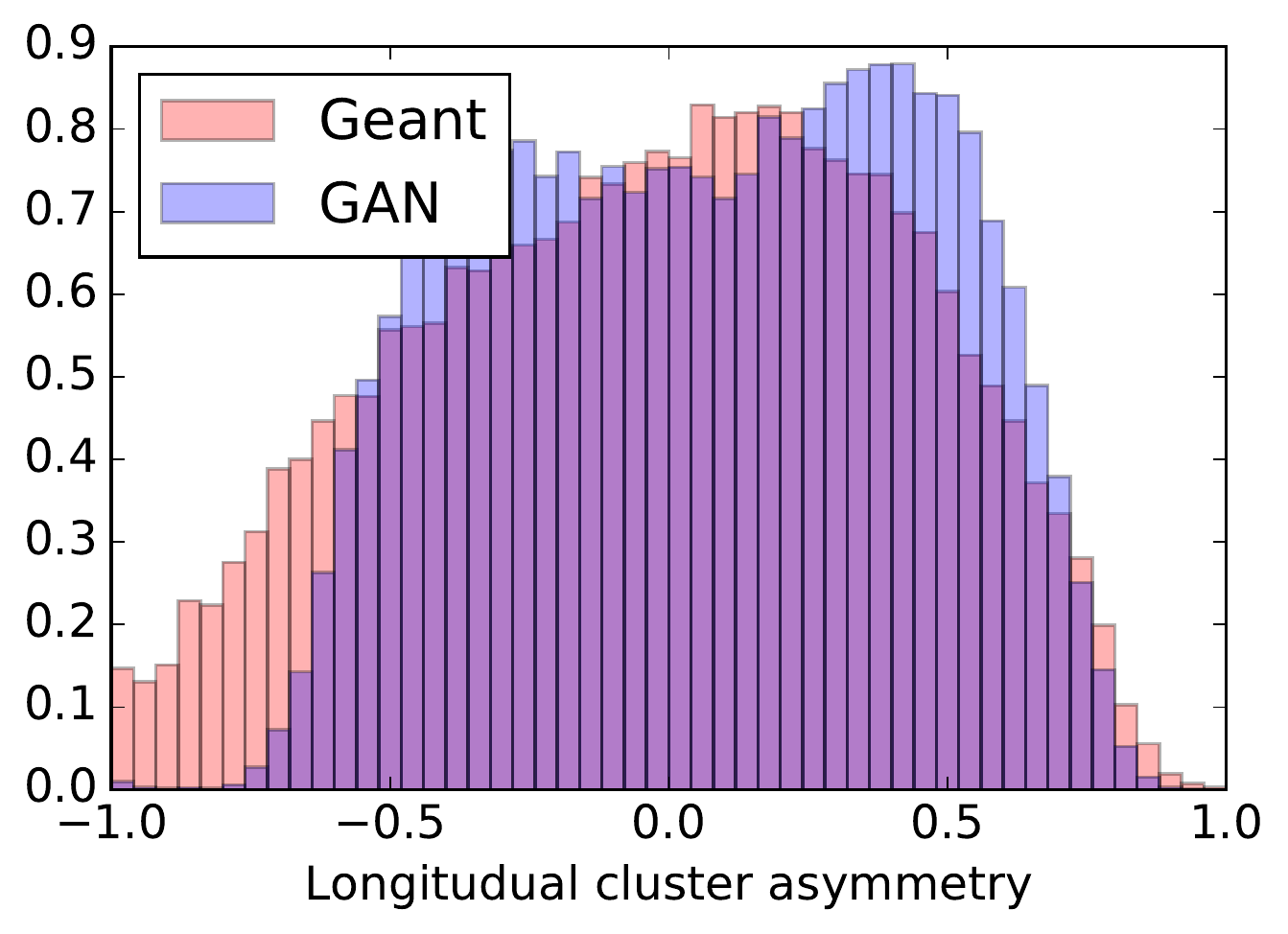}
    \caption{The longitudinal asymmetry of real and generated clusters}
  \end{subfigure}
  \caption{Generated images quality evaluation including described physical characteristics.}\label{fig:quality}  
\end{figure}

Then we continue with a quantitative evaluation of the proposed simulation
method. While generic evaluation methods for generative models exist,
here we base our evaluation on physics-driven similarity
metrics. These metrics are designed using the domain knowledge and the
recommendations from physicists on the evaluation of simulation
procedures. 
For this presentation, we selected a few cluster properties which essentially
drive cluster properties used in the reconstruction of calorimeter objects
and following physics analysis. If the initial particle direction is not
perpendicular to the calorimeter face, the produced cluster is elongated
in that direction. Therefore, we consider separately cluster width in
the direction of the initial particle and in the transverse
direction. Spatial resolution, which is the distance between the centre
mass of the cluster and the initial track projection to the shower max
depth, is another important characteristic affecting the physics
properties of the cluster. Cluster sparsity, which is the fraction of
cells with energies above some threshold, reflects the marginal low
energy properties of the generated clusters. Finally, longitudinal and
transverse asymmetries, which are differences in energies between
forward-backwards and left-right sides of the cluster, characterise
coherent energy variations.  
 A comparison of these characteristics is presented in~\cref{fig:quality}. 

The primary cluster characteristics demonstrate good agreement with
fully simulated data. However, secondary characteristics driven by
long-range correlations between different cluster contributions might
be significantly improved.

As for model performance, we trained our model for 3000 epochs which take about 70 hours on GPU NVIDIA Tesla K80. The sampling rate is 0.07 ms per sample on GPU, 4.9 ms per sample on CPU.

\section{Conclusion and outlook}\label{conclusion}

The research proves that Generative Adversarial Networks are a good candidate for fast simulation of high granularity detectors typically studied for the next generation accelerators. We have successfully generated images of shower energy deposition with a condition on the particle parameters, such as the momentum and the coordinates, using modern generative deep neural network techniques such as Wasserstein GAN with gradient penalty.

Future work will be focused on improving reproduction of second-order cluster characteristics, such as variations and long-range correlations between different cells.

The research leading to these results has received funding from the Russian Science Foundation under agreement No 19-71-30020.

\bibliography{caloGAN_chep2018}

%\appendix
%\input{appendix.tex}

\end{document}